\documentclass[conference]{IEEEtran}

\usepackage[utf8]{inputenc}
\usepackage{balance}
\usepackage{xcolor}
\usepackage{graphicx}
\usepackage{float}
\usepackage{csquotes}
\usepackage{tabularx}
\usepackage{framed}
\usepackage{fancybox}
\usepackage{url}
\usepackage{multirow}
\usepackage{booktabs}
\usepackage{tikz}
\usepackage{comment}
\usepackage{cite}

\newcommand\copyrighttext{%
      \footnotesize \textcopyright~2024 IEEE. This document is a preprint. Personal use of this material is permitted.
      Permission from IEEE must be obtained for all other uses, in any current or future
      media, including reprinting/republishing this material for advertising or promotional
      purposes, creating new collective works, for resale or redistribution to servers or
      lists, or reuse of any copyrighted component of this work in other works. DOI: will be provided once available.
      }
    \newcommand\copyrightnotice{%
    \begin{tikzpicture}[remember picture,overlay]
    \node[anchor=north,yshift=-12pt] at (current page.north) {\fbox{\parbox{\dimexpr\textwidth-\fboxsep-\fboxrule\relax}{\copyrighttext}}};
    \end{tikzpicture}%
}

\title{Explanations in Everyday Software Systems: Towards a Taxonomy for Explainability Needs}

\makeatletter % changes the catcode of @ to 11
\newcommand{\linebreakand}{%
  \end{@IEEEauthorhalign}
  \hfill\mbox{}\par
  \mbox{}\hfill\begin{@IEEEauthorhalign}
}
\makeatother % changes the catcode of @ back to 12

\author{\IEEEauthorblockN{Jakob Droste}
\IEEEauthorblockA{\textit{Leibniz Universität Hannover}\\
\textit{Software Engineering Group} \\
Hannover, Germany \\
jakob.droste@inf.uni-hannover.de}
\and
\and
\IEEEauthorblockN{Hannah Deters}
\IEEEauthorblockA{\textit{Leibniz Universität Hannover}\\
\textit{Software Engineering Group} \\
Hannover, Germany \\
hannah.deters@inf.uni-hannover.de}
\and
\IEEEauthorblockN{Martin Obaidi}
\IEEEauthorblockA{\textit{Leibniz Universität Hannover}\\
\textit{Software Engineering Group} \\
Hannover, Germany \\
martin.obaidi@inf.uni-hannover.de}
\and
\linebreakand % <----- NOTE HERE, breaking after the third one!
\IEEEauthorblockN{Kurt Schneider}
\IEEEauthorblockA{\textit{Leibniz Universität Hannover}\\
\textit{Software Engineering Group} \\
Hannover, Germany \\
kurt.schneider@inf.uni-hannover.de}
}

\begin{document}
\maketitle
%\IEEEpeerreviewmaketitle

\copyrightnotice
\vspace{-2ex}

\begin{abstract}
% Context and motivation
Modern software systems are becoming increasingly complex and opaque. The integration of explanations within software has shown the potential to address this opacity and can make the system more understandable to end-users. As a result, explainability has gained much traction as a non-functional requirement of complex systems.

% Question/problem
Understanding what type of system requires what types of explanations is necessary to facilitate the inclusion of explainability in early software design processes. In order to specify explainability requirements, an explainability taxonomy that applies to a variety of different software types is needed. 
% Principal ideas/results
In this paper, we present the results of an online survey with 84 participants. We asked the participants to state their questions and confusions concerning their three most recently used software systems and elicited both explicit and implicit explainability needs from their statements. These needs were coded by three researchers. In total, we identified and classified 315 explainability needs from the survey answers.

% Contribution
Drawing from a large pool of explainability needs and our coding procedure, we present two major contributions of this work: 1) a taxonomy for explainability needs in everyday software systems and 2) an overview of how the need for explanations differs between different types of software systems.
\end{abstract} 

\begin{IEEEkeywords}
Requirements Engineering, Explainability, Taxonomy, User Feedback
\end{IEEEkeywords}

\section{Introduction}
\label{sec:intro}

The development of software systems is becoming increasingly complex~\cite{kraut1995complex,perry1994complex}. This is reflected in the fact that development is carried out by teams of developers, some of whom are distributed globally~\cite{kuhrmann2018helena}. Consequently, developers are confronted with the challenge of designing and implementing software solutions that not only meet functional requirements but also adhere to a multitude of non-functional requirements (NFRs) like security or usability.
A relatively new non-functional requirement is explainability~\cite{chazette2020explain,chazette2022slr,koehl2019explainnfr}. Explainability usually aims to make the decisions and results of algorithms transparent and understandable, especially in highly complex applications, such as in the fields of artificial intelligence (AI) and machine learning~\cite{confalonieri2020explAI,angelov2021eplAI,goodwin2022explML,burkart2021survey}. But even beyond machine learning, there are areas where users need or demand an explanation, such as in relation to the quality aspect of privacy~\cite{brunotte2023PrivacyExplanations}. For example, if an app wants to access the smartphone's contacts, many users want an explanation of why the app needs this access.

%\textit{Problem Statement.}
Explanations should be integrated carefully, as a random insertion of explanations at any point can have a negative effect on the user experience~\cite{chazette2020explain}. An understanding of the specific types of explanations required by different users in different software contexts is essential. %This is the prerequisite for specifying explainability requirements and providing explanations for these requirements in the software development process. However, the lack of an existing taxonomy makes it difficult to accurately categorize explanation needs. 
It has been found that the elicitation of requirements is supported by having tools like quality models or checklists that can be used as a guide~\cite{doerr2005NFRinIndustry}. For example, the quality model ISO 25010 is an established tool for identifying non-functional requirements in the elicitation process. To simplify the elicitation of specific explanation requirements, a reference is needed that describes what kind of explanation needs exist.

A taxonomy that contains the different types of explanation needs can serve as a checklist, giving the requirements engineers guidance to discuss with the customer or users which explanations are desired~\cite{lauenroth2017TaxonomiesInRE}. In addition, a taxonomy offers a clearly defined terminology which helps to express explanation requirements in the further requirements engineering process. In order to create such a taxonomy, however, it is necessary to find out which types of explanation requirements actually exist in different software systems.

In this paper, we develop a taxonomy to categorize explanation needs of users facing everyday software systems. To achieve this goal, we conducted a online study with 84 participants and asked them about their three most recently used softwares. Based on their feedback, we identified explanation needs and categorised them throughout multiple labeling rounds.

Through our study and analysis we provide two main contributions:
\begin{enumerate}
    \item an approach for identifying the need for explanation. 
    \item a taxonomy that can be used to categorize explanatory needs. 
\end{enumerate}
These contributions can help requirements analysts to specify explainability requirements in the development process.

The rest of the paper is structured as follows: In Section \ref{sec:backgroundandrw}, we present related work and background details. The study design is introduced in Section \ref{sec:studydesign}. Section \ref{sec:results} summaries the results that are discussed in Section \ref{sec:discussion}, before concluding the paper in Section \ref{sec:conclusion}.

\section{Background and Related Work}
\label{sec:backgroundandrw}

\subsection{Explainability}
Explainability is a non-functional requirement (NFR) that is often considered in the context of AI systems~\cite{chazette2020explain}. However, systems without AI are also becoming increasingly complex, which is why explainability is also considered outside the explainable artificial intelligence (XAI) domain.
Simplifying the definition by Chazette et al.~\cite{chazette2021ExploringExplainability}, a system is described as explainable with respect to an aspect X if a corpus of information (the explanation) is given to an addressee A which enables A to understand X. Chazette et al.~\cite{chazette2021ExploringExplainability} do not define which aspects X should be explained, but give examples such as the behavior of the system or knowledge about the user. XAI methods focus on explaining the underlying model, more precisely the internal operations to justify decisions made~\cite{das2020XAISurvey}. Brunotte et al.~\cite{brunotte2023PrivacyExplanations} emphasize the importance of privacy explanations that reveal the purpose of using personal information. Besides algorithms and privacy, Deters et al.~\cite{deters2023ExplanationsOnDemand} also mention information regarding interactions with the system as a possible aspect that should be explained. Interaction explanation are intended to guide the user in using complex systems. 

Explainability has an impact on many other NFRs. For example, explainability is often regarded as a means to increase trust in a system~\cite{das2020XAISurvey,rossi2018XaiTrust,shin2021XAITrustAndAcceptance}. However, it was also found that explanations do not necessarily facilitate trust~\cite{kaestner2021ExplainabiltyNotTrust}. Likewise, explanations can have both a positive and a negative effect on usability and understandability~\cite{chazette2021ExploringExplainability}. The effect on transparency and learnability was found to be mainly positive~\cite{chazette2021ExploringExplainability}. Different types of explanations can therefore have different impacts on other NFRs.

\subsection{Taxonomies}
A taxonomy is a classification system that is mostly used for animals and plants, but is also applied in the field of software engineering~\cite{TaxonomyMappingStudy2017}. Classifications can help to identify gaps in a knowledge field, and they can enable a better understanding of the connections between objects~\cite{ClassificationsInSE2009}. Hierarchical taxonomies consist of one top class with several sub-classes which can also have further sub-classes. A true hierarchy ensures mutual exclusivity, which means that an entity belongs to exactly one class. This type of taxonomy is the most common in the software engineering sector~\cite{TaxonomyMappingStudy2017}. Taxonomies in software engineering are most frequently developed for the knowledge areas of software construction, software design, software requirements and software maintenance~\cite{TaxonomyMappingStudy2017}. Since explainability is considered an NFR, our taxonomy belongs to the group of software requirements.

\subsection{Related Work}
Speith~\cite{speith2022XAITaxonomies} conducted a review of taxonomies for XAI methods. He identified eleven representative papers that reference or propose a taxonomy for explainability methods. 
Ferreira et al.~\cite{ferreira2020ExplainabilityTaxonomy} developed a general taxonomy for explainability using an SLR. They describe three aspects of explainability. Firstly, they describe the reasons why explainability is introduced, secondly, they describe who receives explanations and thirdly, they describe the domain in which explainability is integrated. 
There are also several other taxonomies for designing and evaluating explainability~\cite{sokol2020MetricTaxonomy, doshi-velez2017MetricTaxonomy,inunes2017DesigningTaxonomy}. 
%The large number of existing taxonomies shows that they are a common tool in the field of explainability. 
However, all the above mentioned taxonomies do not refer to what types of explanations users need.

Unterbusch et al.~\cite{ExolainabilityUnterbusch23} created a taxonomy for the need for explanation by analyzing 1730 app reviews of eight smartphone apps. They divide the need for explanation into primary and secondary concern. By primary concern, they mean that the need for explanation is the user's only issue. Secondary concern means that there is another issue that could be explained to increase the understanding of the situation, but cannot be solved by an explanation. For example, an error cannot necessarily be solved by an explanation, but understanding the error can reduce frustration. Using app reviews as the foundation for the taxonomy has two limitations that we try to overcome with our methodology. Firstly, the evaluation is limited to smartphone apps, which may have a different need for explanation than general software systems. Secondly, the threshold for writing an app review is quite high, which makes a bias in the reported problems likely. That means that mainly large problems, which create a high level of frustration that leads to writing a review, were considered. With our methodology, we try to cover everyday software systems and deal with all kinds of problems that lead to a need for explanation.

\section{Study Design}
\label{sec:studydesign}

\subsection{Research Questions}

%... we strive towards reaching the following goal formulated as proposed by Wohlin et al. \cite{wohlin2012experimentation}:\\

%\noindent \fbox{%
%\noindent \parbox{\dimexpr\linewidth-2\fboxsep-2\fboxrule}{%
%\underline{\textbf{Research Goal: }}\\
%\textit{Analyze} user feedback 
%\textit{for the purpose of} identifying explanation need
%\textit{with respect to} software
%\textit{from the point of view of} a requirement engineer
%\textit{in the context of} survey study.}}

The main objective of our research is to identify and classify explanation needs of users regarding everyday software systems. Accordingly, our research in this work was framed by two research questions:

\begin{enumerate}
    \item[RQ1] What types of needs for explanations do end-users have in everyday software systems?
    \item[RQ2] How does the need for explanation differ between different types of software systems?
\end{enumerate}

To answer \textit{RQ1}, we conduct our online survey and process the resulting data set to build our taxonomy for explainability needs. We then apply the taxonomy to the same data set to examine differences between software types and answer \textit{RQ2}.

\subsection{Instrument Development}
%We used the survey method~\cite{robson2016real} to collect our data, implemented as an online questionnaire. 
We conducted a qualitative survey in the form of an online questionnaire to collect a comprehensive set of data. Earlier studies on explanation needs observed that explicit questions of whether explanations are desired in specific cases may lead to an affirmation bias~\cite{droste2023personas}. In order to elicit the needs for explanations without asking suggestively, we designed open-ended questions aimed at identifying needs for explanation that emerged in past uses. These questions allowed the participants to put themselves in a real context and thus capture their actual needs for explanation. To test the comprehensibility and effectiveness of the questionnaire, we conducted a pilot study with 8 participants. As a result, some questions were revised to improve their comprehensibility, whereas the structure of the questionnaire remained the same.

\subsection{Survey Structure}

The survey can be divided into three sections. In the first section, participants were asked to indicate which three software systems they had used most recently. In the second section, the participants were then asked six questions about each of the software systems they indicated. The first question was designed to get participants to think back to when they used the system. The remaining five questions were aimed at determining at which points explanations were needed. The questions were posed in different ways in order to activate the participants' memories. Lastly, the third section captured demographics. The entire questionnaire is included in our supplementary material~\cite{SupMat}.

\subsection{Data Collection}
We offered the survey in both German and English via LimeSurvey\footnote{\url{https://www.limesurvey.org/}}. The survey was open from January 2023 to March 2023 and was distributed via networking platforms (like LinkedIn) and the institution's bulletin board. Since our target group was all adults with access to technology and there were no other requirements for participation, convenience sampling was appropriate. 

Three types of data were collected - firstly, demographic data such as gender, age and field of work. Secondly, software systems that were most recently used were collected - this data consists of free text responses, usually consisting of one or two words. Finally, we asked users about various aspects of using their software. For example, what frustrated them about using the software, what confused them, whether they had or have questions about the software and whether they would like to know anything else about the software. By asking these questions, we hope to be able to indirectly elicit the need for explanations from users. This data was collected in the form of free text answers consisting of complete sentences or bullet points.
%- wie verbreitet, welche Teilnehmer, Zielgruppe, was gefragt, Umfragestruktur
%- Verteilungszeitraum
%- Limesurvey
%- In excel tabelle

\subsection{Demographics}
A total of 84 participants completed the survey. Surveys that were not fully completed were removed from the data set. This was done to ensure that answers to certain parts of the questionnaire would not outweigh others throughout the later coding procedure.
35\% of the participants were female, 64\% were male and 1\% was diverse. The average age was 37,8  (min: 18 years, max: 72 years, SD: 16,3). 17\% of participants were studying, 55\% were working and 14\% were both studying and working. 14\% of the participants stated that they were neither studying nor working.

\subsection{Data Analysis}

\begin{figure}[t]
\includegraphics{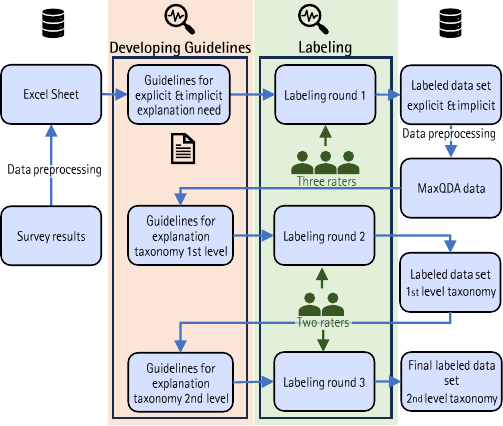}
\caption{Approach of our data analysis}
\label{fig:approach}
\end{figure}

An overview of our data analysis approach can be seen in Figure \ref{fig:approach}.
We exported the results of our survey via LimeSurvey into an Excel spreadsheet.
Three of the authors of this paper then pre-processed the survey answers for the need for explanation. The three raters labeled each of the participants' responses that could contained needs for explanation. The label categories were: explicit explanation need, implicit explanation need, unspecific explanation need and no explanation need. 

Explicit explanation need involves the explicit statement of a need for explanation. An example would be if a user gave the following answer when asked if he or she had any questions about a software: ``Yes, how can I quote a group message and then reply to it?''.
An implicit explanation need, on the other hand, can be recognized by certain trigger words, e.g. when something is not obvious to the participant, is not comprehensible or seems questionable. This could be, for example, the following feedback of an user regarding a software: ``Some functions are not immediately obvious and you need to follow specific instructions''. 
Unspecific explanation need refers to explanation need where a participant responds that he or she has some kind of explanation need without specifying what the need is.
In order to be as objective as possible when labeling according to these categories, we have developed guidelines containing indicators and examples for the four labeling categories. These guidelines were derived by labeling the first five participants' responses in the data set and then discussing the labeling results.

The three raters then labeled the entire data set in two cycles (50\% each). 
In order to prepare for the second labeling round, we also tried to find suitable categories during the first labeling round. Therefore, during the first labeling round, the three raters noted possible categories for the identified explanation needs.
After labeling, the raters reviewed the categories and developed a taxonomy based on these categories. Using this taxonomy, the raters reanalyzed the first five data points, discussed the labeling results and adjusted the taxonomy accordingly. We used Excel for the first labeling round.

\begin{figure*}[t]
    \centering
    \includegraphics[width=.77\linewidth]{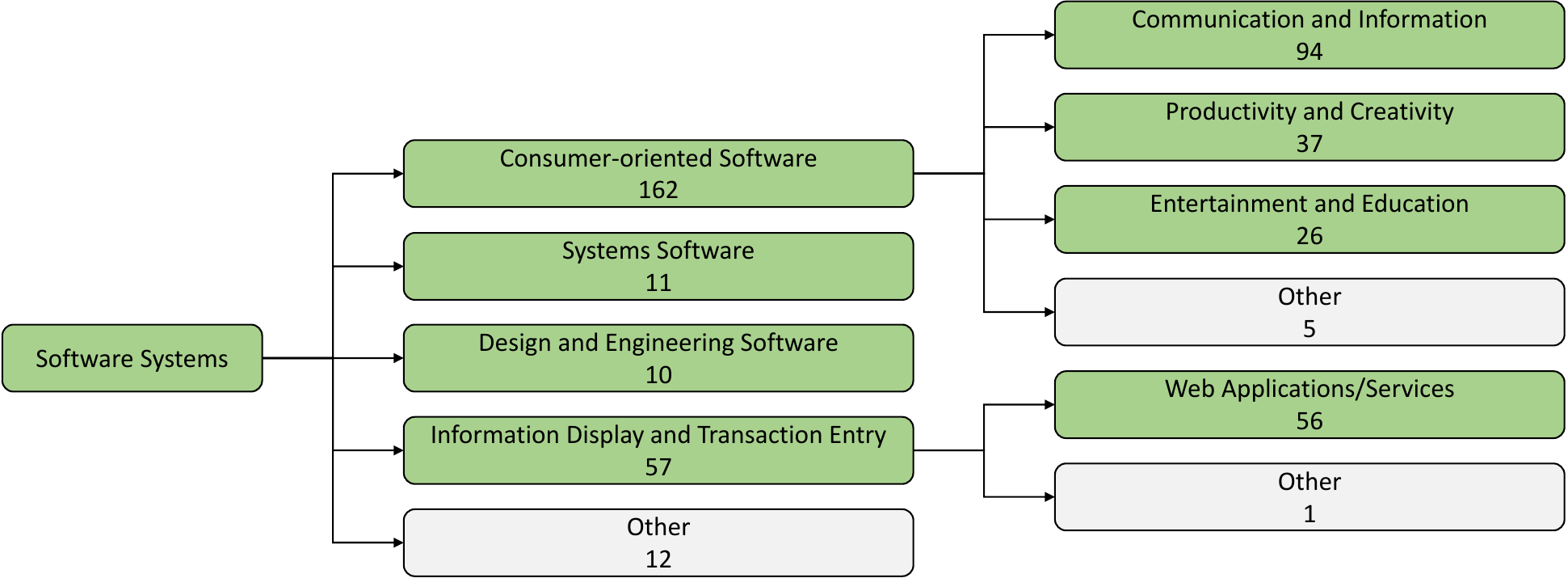}
    \caption{Software types examined in our study (classified according to Forward and Lethbridge~\cite{forward2008taxonomy})}
    \label{fig:types}
\end{figure*}

The entire data set was then re-labeled in two cycles (again 50\% in each cycle), this time by two raters. If there was disagreement between the two, the third rater was consulted in order to reach an agreement. During this second labeling process, the raters wrote down possible subcategories of the taxonomy categories from the previous labeling round.
Finally, the two raters re-labeled these subcategories based on the first five participant responses. In the end, the complete data set was re-categorized into these subcategories in one cycle by two raters. Again, the third rater was consulted in order to reach an agreement in case of disagreement. We used the coding software MAXQDA\footnote{https://www.maxqda.com/} for second and third labeling round.

Finally, all software systems mentioned by the participants were categorized into software type categories by two raters according to an existing taxonomy for software types~\cite{forward2008taxonomy}. There were no disagreement between the two raters.

\subsection{Interrater Agreement}
After each labeling step, we resolved any disagreement between raters. We used two methods to measure disagreement.
To measure the extent to which the three raters agree in their labeling results during the first labeling round, we calculated \textit{Fleiss' Kappa} ($\kappa$) \cite{fleiss1971measuring} as agreement value. Fleiss' $\kappa$ measures the interrater agreement between more than two raters \cite{fleiss1971measuring}. 
For the other labeling rounds, where two raters labeled, we calculated Brennan \& Prediger $\kappa$ (B \& P $\kappa$) \cite{brennan81kappa}, an adapted kappa from Cohen's $\kappa$ \cite{cohen60kappa}.

We classify the $\kappa$ value according to Landis and Koch~\cite{landis1977measurement}.
Furthermore, we calculated the proportional agreement between the raters during the first labeling round. That is the proportion between the agreement of the three raters and the total amount of data to be labeled. For example, if the raters labeled a data set and agreed 70 times on the label and disagreed 30 times, the proportional agreement between them would be 70\%.

\subsection{Data Availability Statement}
To enable the verification and replication of our work, we provide a supplementary material~\cite{SupMat}. In this material, the questionnaire and coding guidelines used in our study are available. Furthermore, the material includes a CSV-file that contains all software systems mentioned by the participants of our study with the corresponding system type and explainability need codes, including categories and subcategories. Participants' uncoded survey answers cannot be shared openly, due to privacy concerns. However, they are available from the first author of this paper on reasonable request. All samples of uncoded survey answers found within this paper have been adapted to ensure the anonymity of our participants.

\section{Results}
\label{sec:results}
\subsection{Types of Software Systems}
Participants answered the questionnaire with regard to their three most recently used software systems. In accordance with our sample size of 84 respondents, we collected 252 names of software systems. Notably, the same application could be mentioned by multiple participants. We labeled the reported systems in accordance with the software type taxonomy proposed by Forward and Lethbridge~\cite{forward2008taxonomy}. An overview of the software types can be seen in Figure~\ref{fig:types}. Due to the large number of different software types, we classified software types that appeared less than 10 times as ``other'', to increase the readability of this paper. The detailed classification of each individual software named by our participants can be found in our supplementary material~\cite{SupMat}.

The participants named systems that they use in their daily lives. Unsurprisingly, the majority of reported systems were consumer-oriented software~(162). This category includes communication and information software~(94) such as messenger apps and e-mail programs, software for productivity and creativity~(37) such as office applications, and entertainment and education software~(26) like media streaming services. 
The second most prevalent software type was applications for information display and transaction entry~(57). Examples of these are web applications~(56) such as social networks, e-finance and user-generated content like image boards. The remaining prominent categories were systems software~(11) such as kernels, password managers and anti-virus applications, and design and engineering software~(10) like video and music composition tools.

\subsection{Taxonomy for Explainability Needs}
The main contribution of this work is the taxonomy that we developed as a result of our coding procedure. We report our interrater agreement in Table~\ref{tab:interrater}.

\begin{table}[t]
\caption{Interrater agreement values during labeling}
\label{tab:interrater}
\begin{tabular}{@{}lrrrrrrr@{}}
\toprule
                             & \multicolumn{3}{r}{Round 1} & \multicolumn{3}{r}{Round 2} & Round 3 \\ \midrule
\% Data                      & 0-50    & 50-100   & 100    & 0-50    & 50-100   & 100    & 100     \\
\# Raters                    & 3       &3         & 3      & 2       & 2        & 2      & 2       \\ \midrule
Fleiss $\kappa$              & 0.87    & 0.81     & 0.84   &         &          &        &         \\
B \& P $\kappa$ &         &          &        & 0.81    & 0.74     & 0.77   & 0.76    \\
Proportion            & 0.87    & 0.81     & 0.84   & 0.83    & 0.76     & 0.79   & 0.77    \\ \bottomrule
\end{tabular}
\end{table}

The $\kappa$ values are always above 0.70, often even above 0.80. This means that we interpret the kappa values as substantial to almost perfect agreement~\cite{landis1977measurement}. The proportional agreement values are also over 75\%, which show that in most of the time the two or three authors agreed.

\begin{figure}[t]
    \centering
    \includegraphics[width=\linewidth]{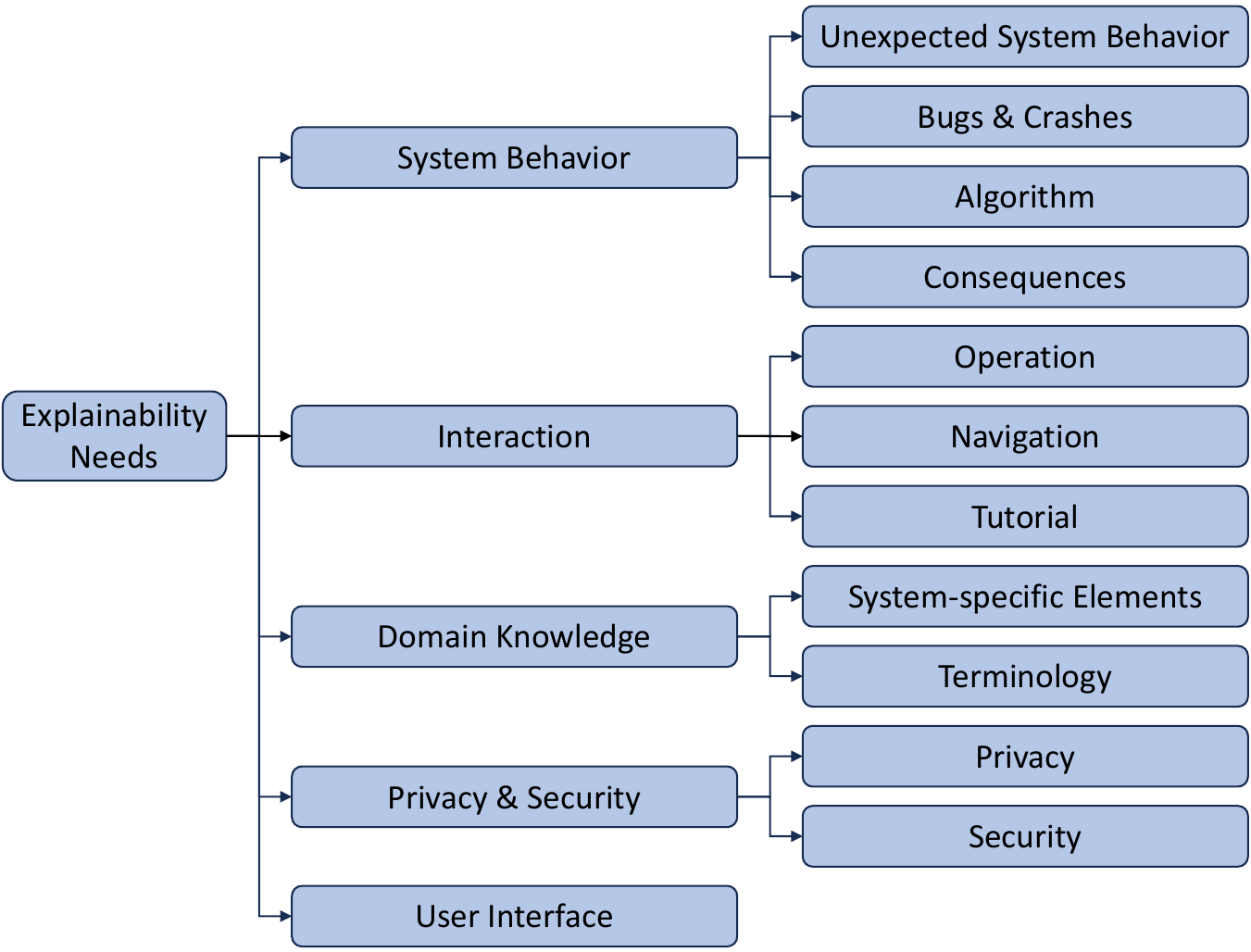}
    \caption{Taxonomy of explainability needs in everyday software systems}
    \label{fig:taxonomy}
\end{figure}

We found five categories of explainability needs within our data set, which we divided into 11 subcategories. These categories and subcategories are displayed in Figure~\ref{fig:taxonomy}. We were able to assign each explicit and implicit explainability need to at least one subcategory. Examples of each kind of explainability need are provided in Table~\ref{tab:needsacross}. The classification of each individual need and the coding guidelines for each category can be found in our supplementary material~\cite{SupMat}.

\subsubsection{System Behavior}
The first category is the need to explain \textit{System Behavior}. This covers explaining how the software works and why it behaves in certain ways. \textit{Unexpected System Behavior} can be a major cause of confusion and frustration for end-users. This need for explanation arises when the observed system behavior deviates from what the user expects. In specific cases, users may directly relate unwanted system behavior to \textit{Bugs \& Crashes}. Explanations can help users understand these errors and enable them to work on a solution. On other occasions, users might wonder about the \textit{Consequences} of their inputs, i. e., how the system will respond if they perform a certain action. Commonly used in XAI are explanations of an \textit{Algorithm} and its outputs. However, these explanations may also be required for complex algorithms outside of AI. Explaining how and why a software behaves in certain ways could increase the understandability and transparency of the system.

\subsubsection{Interaction}
The second category is the need to explain \textit{Interactions} between the end-user and the system. Specifically, users want to know how a certain \textit{Operation} with the software can be performed. Operations may concern all kinds of software functionalities, such as the user making inputs or ordering the system to perform tasks. Operation explanations are differentiated from explanatory needs concerning the \textit{Navigation} within the software. Instead of specific operations on software functionalities, users want to learn how to reach and access different views or subsystems within a system. In other cases, where users are concerned with learning how to use a new system or new features from scratch rather than understanding a specific operation, we categorize the explanatory need as a need for a \textit{Tutorial}. Explaining how a system can be used and navigated has the potential to increase the usability and learnability of the system.

\begin{table*}[t]
    \centering
    \caption{Distribution of explainability needs across all software types and example needs}
    \label{tab:needsacross}
    \begin{tabular}{@{}lrrl@{}}
    \toprule
    \textbf{Type of need}        & \textbf{\#}  & \textbf{\%}     & \textbf{Example adapted from the data set} \\ \midrule
    All needs combined           & 315          & 100\%           &                  \\ \midrule
    \textbf{Interaction}         & \textbf{159} & \textbf{50.5\%} & \textbf{}        \\
    Operation                    & 105          & 33.3\%          & ``How can I set filters to automatically assign my e-mails to certain folders?''                 \\
    Navigation                   & 30           & 9.5\%           & ``Sometimes it is unclear where my podcasts are being saved.''                 \\
    Tutorials                    & 24           & 7.6\%           & ``Due to the large number of features, it took me a long time to learn to use it efficiently.''                 \\ \midrule
    \textbf{System behavior}     & \textbf{95}  & \textbf{30.2\%} &         \\
    Unexpected system behavior   & 50           & 15.9\%          & ``I was wondering why I could not access the events of a shared calendar.''                 \\
    Bugs \& crashes              & 24           & 7.6\%           & ``The error messages are not comprehensible.''                 \\
    Algorithm                    & 15           & 4.8\%           & ``I am unsure about the capabilities of the search functionality.''                 \\
    Consequences                 & 6            & 1.9\%           & ``I was wondering if my messages would arrive if the person unblocked me.''                 \\ \midrule
    \textbf{Domain knowledge}    & \textbf{27}  & \textbf{8.6\%}  &         \\
    System-specific elements     & 14           & 4.4\%           & ``Why aren't all videos available through my business account?''                 \\
    Terminology                  & 13           & 4.1\%           & ``The names of settings are unclear. Who simply knows what `experimental sharpening' is?''                 \\ \midrule
    \textbf{Privacy \& security} & \textbf{23}  & \textbf{7.3\%}  &         \\
    Privacy                      & 15           & 4.8\%           & ``It is unclear to me exactly which parts of my personal data are stored.''                 \\
    Security                     & 8            & 2.5\%           & ``How safe is the end-to-end encryption?''                 \\ \midrule
    \textbf{User Interface}      & \textbf{11}  & \textbf{3.5\%}  & ``I got confused after an update, as the positions of some functions were shifted.''       \\ \bottomrule
    \end{tabular}
\end{table*}

\subsubsection{Domain Knowledge}
The third category of explainability needs is concerned with \textit{Domain Knowledge}. Even experienced software users might run into problems if they lack knowledge about the domain of the system. For example, a tech-savvy user might still need help when using tax preparation software. This could be the case if the user needs explanations for certain \textit{Terminology} within the software. Furthermore, explanations for \textit{System-specific Elements} might be required. This is a broad subcategory as it includes all kinds of elements unique to an individual system. Examples could be explaining the privileges of different types of accounts or why specific kinds of information are needed for a registration. Providing explanations concerning domain knowledge could increase end-users' understanding of the software and improve usability.

\subsubsection{Privacy \& Security}
The fourth category contains the need to explain \textit{Privacy \& Security} aspects of a system to end-users. Privacy and security are closely related non-functional requirements. Needs for \textit{Privacy} explanations may include explaining if and how the user's data is stored and processed within a software system. The \textit{Security} subcategory deals with explaining what measures are taken to ensure the security of the data and the system. This covers explaining protective measures such as encryption, but also informing users about vulnerabilities. Explaining aspects of privacy and security increases transparency and has the potential to foster end-user trust.

\subsubsection{User Interface}
The fifth and final category is the need to explain the \textit{User Interface}. Users may wonder why certain design decisions were made within the user interface. Furthermore, changes in the interface, which may be caused by updates or patches, can be a cause for confusion. Explaining how and why an interface changed could potentially increase the usability and learnability of the interface.

\subsection{Explainability Needs in the Data Set}
The prevalence of different types of explainability needs within our taxonomy varied in general, but there were also distinct differences between different software types. Notably, we managed to categorize 100\% of the previously identified implicit and explicit explainability needs using our taxonomy.

\subsubsection{Explainability Needs across all Software Types}
The distribution of the explainability needs across all software types, ordered by numbers of appearance, is shown in Table~\ref{tab:needsacross}.

$50.5\%$ of the explainability needs stated by our participant were categorized as \textit{Interaction} explanation needs. At $33.3\%$, the most ubiquitous subcategory was \textit{Operation} explanations. The other two subcategories appeared much less often, with \textit{Navigation} at $9.5\%$ and \textit{Tutorials} at $7.6\%$.

The need to explain \textit{System Behavior} was the second most prevalent category at $30.2\%$. Here, the most common type of need was the explanation of \textit{Unexpected System Behavior}. Needs to explain \textit{Bugs \& Crashes} were expressed at $7.6\%$, while the need to explain an \textit{Algorithm} or its output appeared only at $4.8\%$. The need to explain \textit{Consequences} was the least common subcategory overall at only $1.9\%$.

The remaining three categories of explanatory needs were much less prevalent than the previous two. Explanations on \textit{Domain Knowledge} made up only $8.6\%$ in total. The subcategories were almost evenly spread, with the need to explain \textit{System-specific Elements} at $4.4\%$ and \textit{Terminology} explanations at $4.1\%$.

Explanations for \textit{Privacy \& Security} aspects made up $7.3\%$ of explainability needs. At $4.8\%$, most of these were \textit{Privacy} concerns. The remaining $2.5\%$ covered the need for \textit{Security} explanations. \textit{User Interface} explanations as a standalone category appeared only at $3.5\%$.

\begin{table}[htb]
\centering
\caption{Relative number of explainability needs\\for each software type}
\label{tab:needsper}
\begin{tabular}{@{}lrr@{}}
\toprule
\textbf{Software type}                       & \textbf{\#Needs / \#Type} & \textbf{Rel. \#Needs} \\ \midrule
\textbf{Consumer-oriented Software}          & \textbf{207/162}        & \textbf{1.28}         \\
Communication and Information                & 118/94                  & 1.26                  \\
Productivity and Creativity                  & 44/37                   & 1.19                  \\
Entertainment and Education                  & 38/26                   & 1.46                  \\ \midrule
\textbf{Info. Display and Transaction Entry} & \textbf{70/57}          & \textbf{1.23}         \\
Web Applications/Services                    & 70/56                   & 1.25                  \\ \midrule
\textbf{Systems Software}                             & \textbf{10/11}                   & \textbf{0.91}                  \\ \midrule
\textbf{Design and Engineering Software}     & \textbf{14/10}          & \textbf{1.4}          \\ \bottomrule
\end{tabular}
\end{table}

\begin{table*}[t]
\centering
\caption{Distribution of explainability needs across different software types.\\(* this also includes subcategories appearing less than 10 times, categorized as ``other''.)}
\label{tab:needsandtypes}
\begin{tabular}{@{}lrrrrrrrrrrr@{}}
\toprule
                                                     & \multicolumn{2}{r}{\begin{tabular}[c]{@{}r@{}}System\\ behavior\end{tabular}} & \multicolumn{2}{r}{Interaction} & \multicolumn{2}{r}{\begin{tabular}[c]{@{}r@{}}Domain\\ Knowledge\end{tabular}} & \multicolumn{2}{r}{\begin{tabular}[c]{@{}r@{}}Privacy \&\\ Security\end{tabular}} & \multicolumn{2}{r}{\begin{tabular}[c]{@{}r@{}}User\\ Interface\end{tabular}} & Total        \\ \midrule
                                                     & \textbf{\#}                         & \textbf{\%}                             & \textbf{\#}   & \textbf{\%}     & \textbf{\#}                          & \textbf{\%}                             & \textbf{\#}                            & \textbf{\%}                              & \textbf{\#}                         & \textbf{\%}                            & \textbf{\#}  \\
\midrule
\textbf{Consumer-oriented Software *}                & \textbf{64}                         & \textbf{30.9\%}                         & \textbf{101}  & \textbf{48.8\%} & \textbf{18}                          & \textbf{8.7\%}                          & \textbf{18}                            & \textbf{8.7\%}                           & \textbf{6}                          & \textbf{2.9\%}                         & \textbf{207} \\
\midrule
Communication and Information                        & 40                                  & 33.9\%                                  & 52            & 44.1\%          & 5                                    & 4.2\%                                   & 16                                     & 13.6\%                                   & 5                                   & 4.2\%                                  & 118          \\
Productivity and Creativity                          & 7                                   & 15.9\%                                  & 33            & 75\%            & 3                                    & 6.8\%                                   & 0                                      & 0\%                                      & 1                                   & 2.2\%                                  & 44           \\
Entertainment and Education                          & 15                                  & 39.5\%                                  & 15            & 39.5\%          & 6                                    & 15.8\%                                  & 2                                      & 5.3\%                                    & 0                                   & 0\%                                    & 38           \\
\midrule
\textbf{Information Display and Transaction Entry *} & \textbf{22}                         & \textbf{31.4\%}                         & \textbf{37}   & \textbf{52.9\%} & \textbf{5}                           & \textbf{7.1\%}                          & \textbf{2}                             & \textbf{2.9\%}                           & \textbf{4}                          & \textbf{5.7\%}                         & \textbf{70}  \\
Web Applications/Services                            & 22                                  & 31.4\%                                  & 37            & 52.9\%          & 5                                    & 7.1\%                                   & 2                                      & 2.9\%                                    & 4                                   & 5.7\%                                  & 70           \\
\midrule
\textbf{Systems Software}                            & \textbf{4}                          & \textbf{40\%}                           & \textbf{2}    & \textbf{20\%}   & \textbf{1}                           & \textbf{10\%}                           & \textbf{3}                             & \textbf{30\%}                            & \textbf{0}                          & \textbf{0\%}                           & \textbf{10}  \\
\midrule
\textbf{Design and Engineering Software}             & \textbf{3}                          & \textbf{21.4\%}                         & \textbf{10}   & \textbf{71,4\%} & \textbf{1}                           & \textbf{7.1\%}                          & \textbf{0}                             & \textbf{0\%}                             & \textbf{0}                          & \textbf{0\%}                           & \textbf{14}  \\ \bottomrule
\end{tabular}
\end{table*}

\subsubsection{Explainability Needs between Software Types}
We found explainability needs for all software types that we examined (see Figure~\ref{fig:types}). The relative number of explainability needs, depending on the software type, is displayed in Table~\ref{tab:needsper}.

At $0.91$ needs per software mentioned, systems software had the lowest number of needs per appearance. That means that when a participant reported about a systems software, we identified $0.91$ explainability needs on average. The next highest value is found in software for communication and creativity, for which we identified $1.19$ needs on average. We found a similar number for web applications and services, as well as software for communication and information, which had an average of $1.25$ and $1.26$ needs respectively. The highest needs were identified for design and engineering software at $1.4$ and for entertainment and education software at $1.46$.

The distribution of the explainability needs between different software types is shown in Table~\ref{tab:needsandtypes}. We do not display software systems categorized as ``other'' individually, as they would hardly be comparable to the defined categories and subcategories. However, we included the ``other'' subcategories in the numbers shown for the overarching categories, i.e., consumer-oriented software and software for information display and transaction entry. In the case of software for information display and transaction entry, this made no difference, as the single data point from the ``other'' subcategories added no new explainability needs.

In this paper, we report the relation of software types to the categories of explanation need, but not the subcategories. This is necessary as the display of more detailed data would be much more space-consuming and hard to read. For more specific results, please refer to the detailed coding data in our supplementary material~\cite{SupMat}. The percentages stated hereinafter refer to the distribution of the different explainability needs within each type of software system.

The need to explain system behavior is prevalent across all software types. Systems software has $40\%$ of its explanatory needs concerned with system behavior. Closely following is entertainment and education software at $39.5\%$. Software for communication and education has $33.9\%$ share of behavior explanation needs, whereas web applications and services come in at $33.4\%$. The category is less prevalent in design and engineering software ($21.4\%$), as well as in software for productivity and creativity (15.9\%).

Interaction explanations, which are the largest category overall, were by far the most prevalent in software for productivity and creativity (75\%), and in design and engineering software (71.4\%). Web applications and services have approximately half ($52.9\%$) of their needs in the interaction category, whereas communication and information software has $44.1\%$ of theirs. Entertainment and education software has $39.5\%$ interaction explanation needs. Systems software only has $20\%$ of its explainability needs concerned with interaction, and is the only type of system where the category does not cover the largest share of needs.

Domain related explanations covered a far smaller share of the explanation needs, also between software types. Their largest share was in software for entertainment and education, where they covered $15.8\%$ of needs. Systems software and software for design and engineering both had a single occasion of domain explanation need, which corresponds to $10\%$ and $7.1\%$ respectively. Web applications and services also had a $7.1\%$ share of their needs in the domain category, but a larger total number of occurrences. Finally, domain explanation needs were less prevalent in software for productivity and creativity at $6.8\%$ and least prevalent in communication and information software at $4.2\%$.

Explanation needs concerning privacy and security did not appear for all software types. More specifically, we found no need in software for productivity and creativity, as well as in design and engineering software. Their largest shares are found in systems software at $30\%$ and in software for communication and information at $13.06\%$. They also appeared in software for entertainment and education at $5.3\%$ and in web applications and services at $2.9\%$.

The smallest category of needs, explanations about the user interface, was only found in three types of software systems. In web applications and services, they covered $5.7\%$ of explanation needs. Software for communication and information had a $4.2\%$ share of interface explanation needs, whereas software for productivity and creativity only had a $2.2\%$ share, which was a single occurrence. Software for entertainment and education, as well as systems software and design and engineering software did not have any need assigned to them.

%\clearpage
\section{Discussion}
\label{sec:discussion}
\subsection{Answering the Research Questions}
\subsubsection{RQ1: What types of needs for explanations do end-users have in everyday software systems?}
Our findings show that end-users have a variety of explainability needs regarding everyday software systems. We found the most prominent category of needs to be the need for \textit{Interaction} explanations. While not typically in the focus of contemporary explainability research, the majority of our participants' concerns and questions were related to issues with the operation or navigation of a software, or our participants expressed the need for a tutorial. Explaining how to operate a system and providing appropriate tutorials could be an effective measure to increase the learnability and usability of a system.

Explaining \textit{System Behavior} is more common in explainability research, specifically in the field of XAI. This covers algorithms and their outputs, but also bugs and crashes or otherwise unexpected system behavior. Furthermore, some of our participants wondered about the possible consequences of inputs they could make. The goal of explaining system behavior is typically to increase the understandability and transparency of a software, and thereby foster trust in end-users.

Besides these two large categories, we also found needs for explanations in the areas of \textit{Domain Knowledge}, \textit{Privacy \& Security} and for \textit{User Interfaces}. Domain-specific terminology or system-specific elements can pose a barrier to entry for users, even if they are generally well-versed in technology. Explaining these terms and elements could mitigate this and enable users to effectively use the systems. Similarly, concerns about privacy and security might push users not to engage with a system, even if the concerns are unwarranted. Providing privacy and security explanations could address these concerns and foster trust in the stakeholders of the system. Confusing or non-intuitive \textit{User Interfaces} can frustrate users and impede usability. However, in the case of complex software systems, this might not always be easily avoidable. In those cases, explaining why an interface is designed in a certain way, or why and how specific elements were updated could help the user learn to use the system.

\subsubsection{RQ2: How does the need for explanation differ between different types of software systems?}
The results of our coding procedure show that there is a notable difference in explainability needs between different types of software systems. For instance, while the need for interaction explanations constitutes the largest category, it is not the most prevalent type of need in systems software and only tied for most prevalent in entertainment and education software. While the sample size for systems software is admittedly low, it makes sense that the interaction explanations are less needed in entertainment and education software, which is supposedly easy to use. 

Indeed, the share of interaction explanation needs is largest in software for productivity and creativity, and in software for design and engineering. This is consistent with these systems offering a variety of tools to work with, some of which might not be intuitive to use. Incidentally, these are also the two types of systems in which the need to explain system behavior is the lowest. This makes sense as most of the system's behavior would be a direct consequence of the user's inputs and the tools they chose to process their inputs.

Software for entertainment and education often comes with system-specific elements like account and subscription models, or different forms of organizing content. Explaining the difference between these models and the terminology behind them makes the software more understandable and transparent to end-users. Accordingly, the need for domain explanations was most prevalent in this type of software. While a share of $10\%$ in systems software is also relatively high, this only accounts for a single need and is therefore caused by the low sample size of system softwares.

Compared to other software types, systems software had a large share of privacy and security explanation needs. This makes sense as systems software also includes anti-virus applications and password managers. Software for communication and information also shows a high need for privacy and security explanations. This includes software such as messenger apps and e-mail systems, through which end-users might store and share sensible, personal information. Privacy and security explanations are needed to provide users with the necessary transparency on how their data is being stored and processed.

With one exception in software for productivity and creativity, all needs for explanations concerning the user interface appeared in either software for communication and information, or in web applications and services. Arguably, this could be caused by both communication software as well as web applications being highly relying on their visual interfaces. However, as this is the least prevalent category of explainability need, it is hard to say whether this conclusion holds any merit or if the observation occurred by chance.

\subsection{Discussion of the Results}
\subsubsection{Applicability of the Taxonomy}
Constructing a taxonomy from empirical data is typically subject to a number of limitations:

\begin{itemize}
    \item the content of the statements made by our study participants is influenced by the questions we asked
    \item the applicability of the taxonomy is limited by the sample of study participants
    \item the applicability of the taxonomy is limited by the context of the study
    \item the robustness of the taxonomy depends on the quality of the coding procedure
\end{itemize}

We designed the questions in our survey as open as possible, and avoided leading questions to influence our participants' answers as little as possible. For example, asking if system behavior had confused or frustrated our participants would likely have led to a larger share of explanation need concerning system behavior. To avoid this, we instead asked our participants if \textit{anything} concerned with the software had frustrated or confused them before.

The demographic distribution of our participants is not entirely proportionate to society at large (two thirds male, one third students). Our target demography was adults with access to technology, from which we were able to cover the most prevalent subgroups in terms of gender identity, employment status and age. Consequently, we are confident that our taxonomy covers the software types and explainability needs that are most relevant to our target demography.

Our taxonomy is mainly applicable to everyday software systems, which was the context of our study. Within the context of everyday systems, our taxonomy is limited by the software types mentioned by our participants. As the responses to our study cover a variety of different software types, we are confident that our taxonomy is applicable with the desired context of everyday software systems. In a time of mobile applications and internet, this notion is underlined by the fact that the most frequently mentioned software types were consumer-oriented software, as well as web applications and services.

Lastly, the taxonomy is only applicable if the underlying coding procedure is well performed and results in sufficiently high interrater agreements. Before performing each of our three labeling rounds, we designed and discussed examples and our guidelines until there was no longer disagreement between the raters. During the first two labeling rounds, we coded the data set in two halves and re-evaluated our guidelines in between, to ensure the robustness of our coding procedure. During the third labeling round, we were confident to label the whole data set at once, without re-evaluating the guidelines in between. This decision was justified by the high interrater agreements that we achieved in the previous two labeling rounds. Throughout our entire coding procedure, our interrater agreements ranged between $0.74$ and $0.87$, which is considered to be substantial to almost perfect agreement~\cite{landis1977measurement}. Considering the large amount of statements to be coded and the number of categories and subcategories, we are confident that our taxonomy is build upon a robust data basis. We attribute these results to our rigorous labeling procedure and detailed coding guidelines.

In summary, our findings show that our taxonomy is applicable to everyday software systems used by adults, and that it could be a valuable tool for requirements engineers that are looking to integrate explainability into everyday software systems.

\subsubsection{Software Types as a Factor in Explainability Engineering}
Whether an explanation is appropriate for a certain scenario depends on a variety of factors. Previous works in explainability research have stated the importance of considering the addressee when providing explanations~\cite{sokol2020one,tintarev2007effective}. In essence, they argue that the characteristics of the end-user determine what kind of explanation is required. Other works have highlighted the importance of the context of use~\cite{brunotte2023context,langer2021we}. Here, the idea is that whether an explanation is appropriate or not depends on what context the software is used in. For example, in safety-critical situations, short and concise explanations might be preferred over long and detailed explanations. Yet another work proposes that the goal of the explainer plays an important role when integrating explanations~\cite{deters2023means}.

To our knowledge, this work is the first to examine different types of software systems as a factor in explainability engineering. While limited to everyday software systems, our results show that explainability needs indeed vary between different software types. Naturally, one would assume that more complex software needs more explanations concerning its behavior, and that more data-driven software comes with a larger need for privacy and security explanations. However, these factors have not been systematically examined through empirical data before.

\subsubsection{Explainability beyond System Behavior}
A large section of explainability focuses on the explanation of system behavior. Specifically, the field of XAI provides a large part of the existing explainability literature. AI systems are typically very complex and often considered to be opaque black-boxes~\cite{adadi2018peeking}. As a result, explainability in XAI is sometimes equated to interpretability~\cite{chazette2021ExploringExplainability}, which describes how well the decisions and outputs of a system can be comprehended~\cite{zhang2021survey}. Following our taxonomy, these are aspects of explaining system behavior, more specifically the XAI algorithms and their outputs.

Our results show that explainability requirements are not limited to AI, but apply to different kinds of software systems with varying degrees of complexity. Furthermore, not all explainability requirements are concerned with the explanation of system behavior. While system behavior needs are a large share of the needs we identified, they are second to the category of interaction explanations. According to our data, explanations for interactions need to move into the focus of requirements engineering for everyday software. Especially in the case of complex interactions, simply providing good usability is not the same as explaining how to operate or navigate a system. A high degree of usability may decrease the need for interaction explanations, but developing a software system that is highly usable for every user group is unrealistic. However, carefully elicited requirements for interaction explanations can support inexperienced users by providing guidance if needed and would not impede the experience of long-time users as long as they are optional.

%Notably, our results apply to everyday software rather than AI. AI systems might indeed be more heavily weighted towards system behavior explanations, although this would need to be examined in future work.

\subsubsection{Addressing Bias in Requirements Elicitation}
Raising explainability requirements in the early development stages is challenging. Without the completed software at hand, stakeholders would have to rely on tacit knowledge to express the need for explanations. Mentally putting oneself into a role or scenario that has not been experienced before may introduce hypothetical bias~\cite{deters2023ExplanationsOnDemand,hypotheticalBiasPlott}, and hinder the requirements elicitation process. Furthermore, asking if explanations are needed in specific cases may lead to affirmation bias~\cite{droste2023personas}, which could wrongfully motivate the implementation of unneeded explanations.

We addressed these challenges by focusing on participants' past experiences with software systems they actually use. We let them set the context of use for themselves and motivated them to report past experiences that typically coincide with explicit and implicit needs for explanations. This way, we were able to raise a variety of explainability requirements that are based on actual user needs, rather than tacit knowledge.

\subsection{Limitations and Threats to Validity}
\label{sec:threats}
In the following, we present threats to validity according to our study. We categorize the threats according to Wohlin et al.~\cite{wohlin2012experimentation} as construct, internal, external, and conclusion validity.

\textit{Construct Validity.}
Categorizing software systems into domain categories might introduce subjectivity. The software types were coded by two authors, and there was no disagreement, but the coding could still be influenced by the raters' perspectives.
The categories developed during the labeling of explainability needs and the subsequent taxonomy may not fully capture the complexity and diversity of explanation needs in software systems. As we achieved overall high interrater agreements, and were able to code all requirements, we are confident that our taxonomy is complete within the context of this work.
As we used a survey as the only method of data collection, our results might be influenced by mono-method bias. In the future, we plan to not only elicit explainability requirements via self-reported needs, but also via other triggers, such as behavioral patterns and physiological triggers.

\textit{Internal Validity.}
The conclusions drawn from the self-reported needs of our participants and the developed taxonomy might be influenced by our coding procedure. Our labeling process, despite guidelines and discussions, and despite our high interrater agreement, may still introduce subjectivity. The interrater agreement measures (Fleiss' Kappa and Brennan \& Prediger Kappa) should be scrutinized for potential disagreements.
Participants were asked to recall their experiences with software systems that they used in the past. Memory inaccuracies or biases may affect the reliability of their responses. We tried to address this by not focusing our study on a specific type of system, but instead having them report about their three most recently used applications.

\textit{Conclusion Validity.}
A sample size of $84$ participants might not be sufficient to cover all explainability needs in the entire area of ``everyday'' software systems. However, every participant reported three systems, making for a total amount of $252$ reported systems, and we found a variety of different software types within this data. With regard to these systems, we identified a total of $315$ explainability needs, which we consider a sufficient number for the purposes of our analysis. Hence, while we cannot claim that our conclusion apply to \textit{all} types of everyday software systems, we are confident that they cover a significant number of system types.

\textit{External Validity.}
We did not perform a hands-on validation of our taxonomy in practice or on other user feedback. Thus, we cannot claim the applicability of the taxonomy in practice. However, as our results are based on participants real experiences, we are confident that our results are robust within the context of the software types we examined.
Convenience sampling was used, and the target group was the entire population. Still, the results may not be generalizable to the broader population, particularly if the sample has specific characteristics. Furthermore, the distribution of participants is not always proportionate (e.g., two thirds are male). The only constraint for participation was that participants had to be adults with access to technology, which we adhered to, but there might still be unknown demographic factors that should be considered. Furthermore, we did not collect any data for software developed specifically for minors (under 18 years old) as part of the study, so our results cannot be generalized for this group of people.
The use of networking platforms and an institution's bulletin board might introduce bias, as certain demographics may be overrepresented or underrepresented. Certain software domains may be underrepresented or not present at all. As this work reports on a diverse variety of software types and due to our high interrater agreements, we are still confident that our results are applicable in the context of everyday software applications.

\section{Conclusion}
\label{sec:conclusion}
%Considering the ongoing rise of AI systems, and the introduction of complex software systems into everyday life, explainability has become an important non-functional requirement of our time. Explaining how a system works and why it behaves in certain ways provides understandability and transparency, and should ultimately foster end-users' trust in the system. Moreover, past works have also shown that explainability is applicable to areas other than AI, for example in the case of privacy explanations. At this point, it is unclear what types of software systems need to what kinds of explanations. A taxonomy of explainability needs could provide requirements engineers with the framework needed to understand what kinds of explanations are required in the systems they are working on.

For this paper, we conducted an exploratory study to identify the explainability needs of end-users in software systems that they actually use in their daily lives. Through an online survey with $84$ participants, we identified $315$ explainability needs in various types of software systems. Through a three-step coding procedure with three raters, we categorized the explainability needs stated by our participants into five categories and $11$ subcategories. These categories and subcategories form the basis for our taxonomy of explainability needs. To examine explainability needs in relation to software type, we categorized the software systems into software types according to an existing taxonomy~\cite{forward2008taxonomy} and applied our own taxonomy to the categorized data set.

Our results show that explainability needs exist for various types of everyday software, not only for AI systems. Furthermore, we identified a variety of explainability needs beyond explaining how a system works or why it arrived at a certain output. Specifically, we find that the need to explain system behavior, which is typically in the focus of XAI research, is not the most prevalent type of explainability need in everyday software systems. Instead, everyday software has its most prevalent explainability need in the need to explain end-users' interactions with the software. Different types of software systems lead to different kinds of explainability requirements. In this light, explainability needs explicit consideration in the requirements engineering process, in order to provide appropriate explanations for each type of software.

As the next step, we plan to validate the applicability of our taxonomy by applying it in a practical setting. To this end, we will integrate the taxonomy into the development process of small-scale software projects. Furthermore, we want to investigate the influence of cultural differences on the need for explanations, as well as the influence of common demographic factors such as age and affinity for technology. Lastly, we will research the elicitation of explainability requirements via triggers such as behavioral patterns and physiological triggers.

\section*{Acknowledgments} 
\label{sec:ack}
\addcontentsline{toc}{section}{Acknowledgments}
This work was funded by the Deutsche Forschungsgemeinschaft (DFG, German Research Foundation) under Grant No.: 470146331, project softXplain (2022-2025).

\bibliographystyle{IEEEtran}
\balance
\bibliography{references}

\end{document}